\documentclass{emulateapj}

\shorttitle{Cosmic-ray acceleration at ultrarelativistic shocks}
\shortauthors{Niemiec, Ostrowski, \& Pohl}
\usepackage{natbib}
\begin{document}

\title{COSMIC RAY ACCELERATION AT ULTRARELATIVISTIC SHOCK WAVES: \\
 EFFECTS OF DOWNSTREAM SHORT-WAVE TURBULENCE}

\author{Jacek Niemiec}
\affil{Department of Physics and Astronomy, Iowa State University, Ames,
IA 50011}
\affil{Instytut Fizyki J\c{a}drowej PAN, ul. Radzikowskiego 152,
 31-342 Krak\'{o}w, Poland}
\email{niemiec@iastate.edu}
\author{Micha\l{} Ostrowski}
\affil{Obserwatorium Astronomiczne, Uniwersytet Jagiello\'{n}ski,
ul. Orla 171, 30-244 Krak\'{o}w, Poland}
\and
\author{Martin Pohl}
\affil{Department of Physics and Astronomy, Iowa State University, Ames,
IA 50011}

\begin{abstract}
The present paper is the last of a series studying the first-order Fermi 
acceleration processes at relativistic shock waves with the method of Monte Carlo
simulations applied to shocks propagating in realistically modeled turbulent
magnetic fields. The model of the background magnetic field structure of 
\citet{nie04,nie06} has been augmented here by a large-amplitude short-wave 
downstream component, imitating that generated by plasma instabilities at the 
shock front. Following \citet{nie06}, we have considered  
ultrarelativistic shocks with the mean magnetic field oriented both oblique and
parallel to the shock normal. For both cases simulations have been performed for 
different choices of magnetic field perturbations, represented by various 
wave power spectra within a wide wavevector range. The results show that the
introduction of the short-wave component downstream of the shock is not
sufficient to produce power-law particle spectra with the ``universal" spectral 
index $4.2$. On the contrary, concave spectra with cutoffs are preferentially 
formed, the curvature and cutoff energy being dependent on the properties of 
turbulence. Our results suggest that the electromagnetic emission observed 
from astrophysical sites with relativistic jets, e.g. active galactic nuclei and 
gamma-ray bursts, is likely
generated by particles accelerated in processes other than the 
widely invoked first-order Fermi mechanism.
\end{abstract}


\section{INTRODUCTION}

Realistic modeling of first-order Fermi cosmic-ray acceleration at 
relativistic shock waves is a difficult task due to the strong dependence of the 
resulting particle spectra on the essentially unknown local conditions at 
such shocks. Therefore, progress in this field mostly arises from the 
increasingly more realistic MHD conditions that are assumed to apply
near the shocks. For mildly relativistic shocks in the {\it test particle}
approach the advances that have been made in the past twenty years involve 
a semianalytic solution of the particle pitch-angle diffusion equation at 
parallel shocks \citep{kir87a}, the effects of different shock compressions in 
parallel shocks \citep{hea88}, the impact of an oblique mean magnetic field for 
subluminal configurations \citep{kir89}, and the effects of oblique 
field configurations at sub- and superluminal shocks considered with an 
increasing degree of complication, applying successively more realistic physical 
approximations \citep{ost91,ost93,bed96,nie04}.
For ultrarelativistic shocks the notion of a ``universal'' spectral 
index was first described by \citet{bed98} and later found in a variety of 
studies \citep{gal99,kir00,ach01,lem03,ell04}, but more recent studies 
\citep{nie06} indicate basic problems with first-order Fermi acceleration, 
in particular in generating wide-energy power-law spectra. The latter work also 
contains a more systematic discussion of the subject literature.

On the other hand, there have been attempts to self-consistently derive the 
electromagnetic shock structure. One very simplified approach applied 
modeling of nonlinear effects in the first-order Fermi acceleration at the 
shock (e.g., Ellison \& Double 2002). The limitations of this technique arise 
from the approximate treatment of particle scattering near the shock and its 
simple scaling with particle energy instead of a realistic accounting for the
microphysics of the process. A more realistic microscopic and self-consistent 
description of the collisionless shock transition may be afforded by 
Particle-In-Cell (PIC) simulations
\citep[e.g.,][]{hos92,sil03,nis03,nis05a,fre04,hed05,jar05}, which follow the 
formation of a relativistic collisionless shock starting with processes acting 
on the plasma scale. In these simulations one typically observes that the 
generation of magnetic and electric fields is accompanied by an evolution of the 
particle distribution function from the background plasma conditions up to 
highly superthermal energy scales. Although providing a wealth of 
information on the collisionless shock structure, the PIC simulations are still 
very much limited in dynamical range, in particular for cosmic-ray particles whose 
energies are many orders of magnitude larger than the plasma particle energies. 

The present paper is a continuation of the studies of \citet{nie04,nie06}. 
Here we attempt to incorporate the results of the PIC simulations concerning
the generation of magnetic fields at relativistic shocks into the test particle 
Monte Carlo simulations of cosmic-ray particle acceleration. As in the 
previous papers, we study ultrarelativistic shock waves propagating in a
medium with a magnetic field that is perturbed over a wide range of 
macroscopic scales. 
However, we also consider an additional short-wave isotropic turbulent 
field component downstream of the shock, analogous to the shock generated 
turbulent fields revealed by the PIC simulations. We treat the amplitude of the
short-wave component as a model parameter and study the effects of the 
small-scale perturbations on the particle spectra and angular distributions 
derived in the previous work.

In what follows, $c=1$ is the speed of light. The integration of particle 
trajectories is performed in the respective local plasma (upstream or downstream) 
rest frame, and we use index ``1" (``2") to label quantities provided in the 
upstream (downstream) frame. 
We consider ultrarelativistic particles with $p=E$. We use dimensionless 
variables, so a particle of unit energy moving in an uniform mean upstream 
magnetic field $B_{0}$ has the unit maximum (for $p_{\perp}=E$) gyroradius 
$r_g(E=1)=1$ and the respective resonance wavevector is $k_{res}(E=1) = 2\pi$.

\section{DETAILS OF THE MODELING}

\subsection{Magnetic Field Structure} \label{bozomath}

As in \citet{nie04,nie06} we assume a relativistic shock wave to propagate 
with velocity {\bf\em u}$_1$ (or the respective Lorentz factor $\gamma_1$) in 
the upstream medium with a
uniform magnetic field {\bf\em B}$_0 \equiv$ {\bf\em B}$_{0,1}$, inclined at an 
angle $\psi_1$ to the shock normal (along {\bf\em u}$_1$ direction), superimposed on
which are isotropic 3D magnetic field perturbations. These perturbations are 
characterized with a power-law wave
power spectrum $F(k)\sim k^{-q}$ that is defined in a wide wavevector range 
($k_{1,min}$, $k_{1,max}$), where $k_{1,min}=0.0001$ and $k_{1,max}=10$. 
Specifically we consider the wave spectral indices
$q=1$, describing the flat power spectrum, 
and $q=5/3$ for the Kolmogorov distribution. 
The integral power of these perturbations is given by the (upstream) 
amplitude 
\begin{equation}
\delta B = \sqrt{\int^{k_{1,max}}_{k_{1,min}} F(k) dk},  
\end{equation}
and $\delta B/B_0$ is one of our model parameters. As described in detail in
\citet{nie04,nie06}, the upstream field perturbations are modeled as the
superposition of static sinusoidal waves of finite amplitude. The downstream 
magnetic field structure, including the turbulent component, is derived as the 
shock-compressed upstream field, and hence the downstream
turbulent magnetic field is naturally anisotropic.  Our method of
applying the hydrodynamical shock jump conditions 
\citep[for the electron-proton plasma,][]{hea88}
preserves the continuity of the magnetic field lines across the shock. 
In what follows, we refer to the field component described above as the 
{\it large-scale} (or {\it long-wave}) {\it background field}. 

The other physical magnetic field component considered in the present study 
is the short-wave turbulence, assumed to be a result of kinetic magnetic
field generation processes acting at the shock. This component, taken for 
simplicity to be isotropic%
\footnote{Theoretical considerations by
\citet{med99} and numerical PIC simulations \citep[e.g.,][]{sil03,nis03,fre04}
show that the relativistic two-stream (Weibel-like) instability at the 
relativistic collisionless shock front leads to the generation of a strong, 
small-scale turbulent field downstream, that is predominantly transversal (2D) 
and lies in the plane of the shock. Because the magnetic field in the nonlinear 
regime of the two-stream instability is sustained by the structure of the ion 
current channels, instabilities in the ion filaments (e.g., kink and/or 
firehose instability) should lead ultimately to the 3D turbulent magnetic field.
}
and static, is imposed upon the nonuniform background magnetic field downstream 
of the shock. The short-wave nonlinear field perturbations are introduced with
a flat spectral distribution in the wavevector range ($10 k_{2,max}$, $100 k_{2,max}$), 
where the shortest downstream waves are the shock-compressed shortest upstream 
waves:
\begin{equation}
k_{2,max} = k_{1,max}\, R\frac{\gamma_1}{\gamma_2}.
\end{equation}
Here $R=u_1/u_2$ is the compression ratio in the shock rest frame, and 
$\gamma_1$ ($\gamma_2$) is the upstream (downstream) shock Lorentz factor. 
The wavevector range of the small-scale turbulent component is chosen
to be one decade in $k$ apart from $k_{2,max}$ to separate the influence of 
these perturbations on the low-energy particle motion from that exerted by
the short-wave component of 
the large-scale background field. This choice also facilitates the use of a 
hybrid method for the calculation of particle trajectories (see \S 2.2). Note also,
that the definition of the short-wave component depends on the Lorentz factor of the 
shock. Below, we refer to the short-wave magnetic field component also as the 
{\it shock-generated turbulence} (``{\it sh}''), to distinguish it from the 
large-scale background field, that exists in the upstream region and is
only compressed upon passage through the shock. 

In our simulations we exclusively study the first-order Fermi acceleration 
process and neglect second-order processes. Therefore, the turbulent magnetic
field components (both short-wave and large-scale background perturbations) can be
considered to be static in the respective plasma rest frames, both upstream and 
downstream of the shock, and electric fields that may exist in the shock 
transition layer are neglected. 

\subsection{Monte Carlo Simulations}

The implementation of short-wave turbulence into the Monte Carlo 
simulations forces us to dispense with the direct integration of the particle 
equations of motion in the analytically modeled
magnetic field \citep[see][]{nie06}. Instead, we resume the derivation
of particle trajectories with the hybrid approach 
proposed in \citet{nie04}. Thus, particle trajectories are directly calculated
from the equations of motion in the large-scale background magnetic field only,
whereas the trajectory perturbations due to the shock-generated small-scale
turbulence are
accounted for through a small-amplitude pitch-angle scattering term. 
Auxiliary simulations have been performed to determine the scattering 
amplitude distributions for various particle energies. In the scattering 
procedure, after each time step $\Delta t$ the
particle momentum direction is perturbed by a small
angle $\Delta\Omega$. The time step itself scales with particle energy 
(or gyroradius), $\Delta t\propto E$, and is chosen so that the condition 
$c\Delta t\gg \lambda_{sh}$ is always fulfilled ($\lambda_{sh}$ is the 
wavelength of shock-generated perturbations), which means that the 
auxiliary simulations follow the particle scattering on the shock-generated 
turbulence well into the diffusive regime. 
Therefore, the scattering amplitude distributions also 
scale linearly with the short-wave turbulence amplitude, and the mean 
scattering angle variations are related to the particle energy as
$\Delta\bar{\Omega}\propto E^{-1/2}$.

An individual simulation run is performed as follows. We start with 
injecting monoenergetic particles (with the initial energy $E_0=0.1$) at random 
positions along the shock front, with their momenta isotropically distributed 
within a cone around the shock normal pointing upstream of the shock. 
In an initial simulation cycle, the injection process is continued 
until the required number of particles, $N$, has been selected---those that 
after being injected upstream and then transmitted downstream of the shock, 
succeed in recrossing the shock front again. Then the calculation of 
individual particle trajectories proceeds through all subsequent 
upstream-downstream cycles. In each cycle, a fraction of the particles escapes 
through a free-escape boundary introduced ``far downstream'' of the shock, 
i.e. at a location from which there is only a negligible chance 
that particles crossing the boundary would return back to the shock. 
To the particles remaining in the simulations the 
trajectory-splitting procedure is applied \citep[see][]{nie04}, 
so the number of particles remains constant in the acceleration process, but 
the statistical weights of the particles are appropriately reduced. 
The final spectra and angular distributions of accelerated 
particles, derived in the shock normal rest
frame for particles crossing the shock front, are averaged over many 
statistically different simulation runs. 

\section{RESULTS}

Because of the limited capabilities of present-day computers, the long-time 
nonlinear development of plasma instabilities leading to the generation of the 
short-wave turbulent downstream magnetic field component cannot yet be fully 
investigated, in particular for electron-ion plasma collision fronts. 
Therefore, the ultimate structure of the small-scale field and the 
global effectiveness of the generation mechanism remain uncertain. 
To estimate the role of the shock-generated 
turbulence in cosmic-ray acceleration at ultrarelativistic shocks, we introduce 
the small-scale wave component as isotropic 3D magnetic field perturbations 
and treat the amplitude of these perturbations as a model parameter.

The spectra of accelerated particles for oblique superluminal shocks are presented 
in Figures \ref{obl1}-\ref{obl4}, and those for parallel shocks are displayed in 
Figures \ref{par10} and \ref{par30}. The amplitude of the short-wave component,
$\delta B_{sh}/\langle B_2\rangle$, as provided in the figures, is measured in 
units of the average downstream perturbed magnetic field strength
$\langle B_2\rangle=\langle(${\bf\em B}$_{0,2}+\delta\!${\bf\em B}$_2)^2\rangle^{1/2}$.
The spectra presented with solid lines has been derived in the turbulence model
without the $\delta B_{sh}$ term \citep[most of them are presented in][]{nie06}.
Some of the energetic particle distributions follow a  power-law
in a certain energy range. In these cases linear fits to the
power-law portions of the spectra are presented, and values of the phase-space
spectral indices, $\alpha$, are given (the equivalent {\it number} spectral index
$\sigma=\alpha -2$). The spectral indices may help to make a
quantitative comparison of the spectra, but care must be exercised,
because the spectral indices depend on the energy range chosen for the fit 
on account of the curvature in most of the spectra. In the case of superluminal
shocks, we use the same low-energy limit for the energy range selected for 
the fits. 

\subsection{Oblique Superluminal Shocks}

\begin{figure}
\includegraphics[angle=0,scale=0.85]{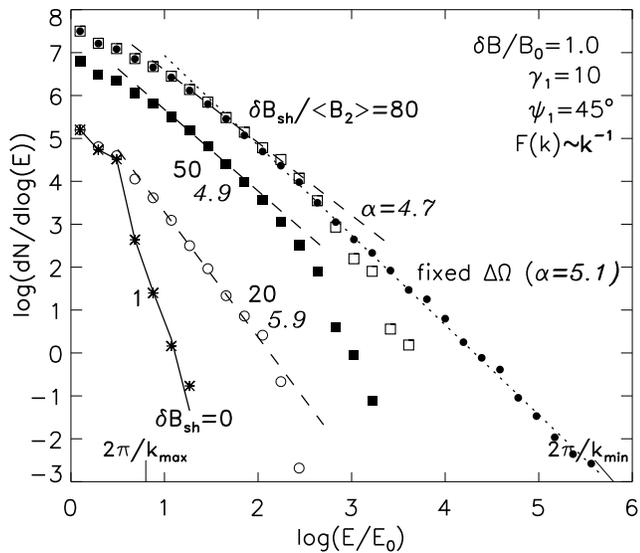}
\caption{Accelerated particle spectra at oblique superluminal shocks with
$\gamma_1 = 10$, $\psi_1 = 45^o$, and a flat wave-power spectrum of 
large-scale background magnetic field turbulence. The amplitude of the
upstream field perturbations is $\delta B/B_{0}=1.0$. The amplitudes of the 
downstream shock-generated turbulence, $\delta B_{sh}/\langle B_2\rangle$, are 
given near the respective spectra. Linear fits to some spectra
are also presented with dashed lines. The energy ranges selected for the fits 
have the same low-energy bounds to facilitate comparison of the spectral
indices $\alpha$, given in italic. 
The spectrum plotted with a solid line applies to the case 
without short-wave shock-generated perturbations. The spectrum shown with
filled dots is calculated with a fixed scattering term for $E\geq 16$, and the
linear fit to the part of this spectrum above $E\geq 16$ is presented with
dotted line. 
Some spectra are vertically shifted for clarity. Particles in the energy range  
($2\pi/k_{max}$, $2\pi/k_{min}$) can satisfy the resonance condition 
$k_{res} = 2 \pi / r_g(E)$ for some of the waves in the background turbulence 
spectrum. 
\label{obl1}}
\end{figure}

\begin{figure}
\includegraphics[angle=0,scale=0.85]{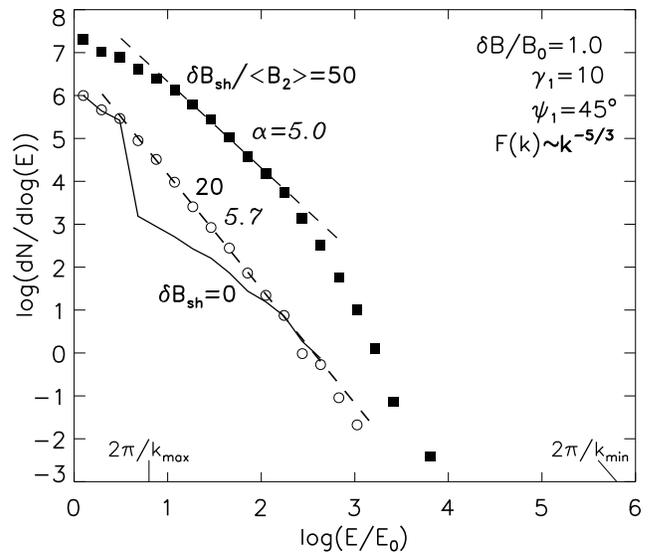}
\caption{Particle spectra derived for oblique superluminal shocks for different
amplitudes of shock-generated turbulence. For all spectra 
$\gamma_1 = 10$, $\psi_1 = 45^o$, and a Kolmogorov power spectrum of
background long-wave turbulence with $\delta B/B_{0}=1.0$ has been assumed. 
The spectrum shown with the solid line is derived in the limit $\delta B_{sh}=0$
\citep[see Fig. 2 in][]{nie06}.
\label{obl2}}
\end{figure}

\begin{figure}
\includegraphics[angle=0,scale=0.85]{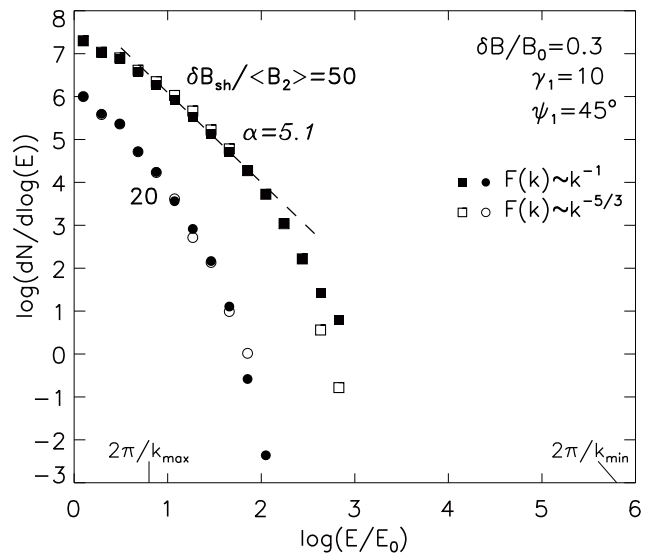}
\caption{Particle spectra at superluminal ($\gamma_1 = 10$, $\psi_1 = 45^o$) 
shocks with $\delta B/B_{0}=0.3$. Filled and open symbols refer to
results for a flat and a Kolmogorov
wave-power spectrum of large-scale background field perturbations, respectively. 
The linear fit to the spectrum for the {\it flat} wave power spectrum is 
derived in an energy range with the same low-energy bound as used for the fits 
shown in Figs. \ref{obl1} and \ref{obl2}. 
\label{obl3}}
\end{figure}

\begin{figure}
\includegraphics[angle=0,scale=0.85]{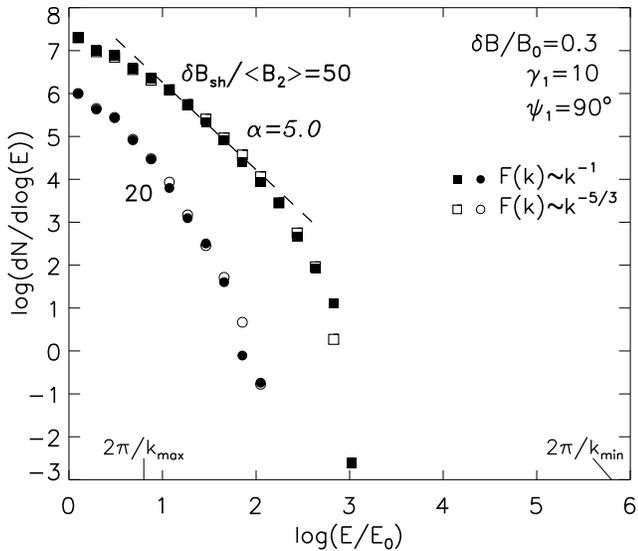}
\caption{Particle spectra at perpendicular ($\gamma_1 = 10$, $\psi_1 = 90^o$)
shocks for the same parameter combinations as in Fig. \ref{obl3}.
\label{obl4}}
\end{figure}

Accelerated particle spectra for oblique superluminal shocks are presented in
Figures \ref{obl1}-\ref{obl3} for $\psi_1=45^o$ ($u_1/\cos{\psi_1} \approx 1.4\, c$),
and in Figure \ref{obl4} for $\psi_1=90^o$ (perpendicular shock).
One can see that increasing the amplitude of the shock-generated turbulence 
leads to a more efficient acceleration with particle spectral tails extending 
to higher energies. However, in all cases, in which  
$\delta B_{sh}/\langle B_2\rangle \gg 1$, the energetic spectral tails are convex, 
so the spectral index increases with particle energy. In addition, all 
the spectra have cutoffs at an energy for which the resonance
condition for interactions with the long-wave turbulence is fulfilled. These 
features result from the fact that the influence on particle trajectories
of the shock-generated small-scale turbulence decreases with increasing particle 
energy. In our numerical approach this corresponds to a reduction
of the scattering amplitude $\Delta\Omega(E)$ (see \S 2.2). 

Most published treatments of first-order Fermi 
acceleration at ultrarelativistic shocks, that apply the pitch-angle diffusion 
approximation \citep{bed98,gal99,kir00,ach01,ell04,kes05}, assume
scattering conditions that do not change with particle energy. As discussed by
\citet{ost02}, these authors show that if the small-angle scattering dominates 
over the possible influence of the oblique mean magnetic
field component and/or the long-wave perturbations in shaping 
the particle trajectories, then power-law particle distributions with 
``universal" spectral index $\alpha_u \approx 4.2$ may be formed.
With the more realistic magnetic field model considered
in the present paper, it is not possible to reproduce these results. However,
particle spectra can approach a power-law form in a limited 
energy range near the particle injection energy, as demonstrated by the
fits to the low-energy parts of the spectra in Figures \ref{obl1}-\ref{obl4}.
The energy ranges for the fits are arbitrary selected but have the same 
low-energy bound. Note that in the figures the spectral indices,
though depending on the background conditions, are {\it always} larger
than $\alpha_u$. If one chose higher energies to derive a power-law fit to
the spectra, one would obtain a higher spectral index on account of
the convex shapes of the spectra. Convex spectra could have been expected
based on the earlier studies of \citet{bed98}, who showed how an increased
pitch-angle scattering angle in superluminal shocks leads to a
flattening of the particle spectrum to the limiting ``universal" power law
\citep[see the discussion of this effect in][]{ost02}. 
To further illustrate this behavior, we have recalculated the spectrum for
$\delta B_{sh}/\langle B_2\rangle=80$ in Figure \ref{obl1} ({\it filled dots})
for a fixed scattering amplitude $\Delta\Omega (E)$ at particle energies
$E\geq 16$ ($\Delta\Omega (E\geq 16)={\rm const.} \ll 1$). Then the pitch-angle 
scattering term dominates downstream of the shock at high particle energies,
which leads to a power-law spectrum formation over a wide energy range without a 
steepening or a cutoff, but with $\alpha > \alpha_u$.

\begin{figure}
\includegraphics[angle=0,scale=0.85]{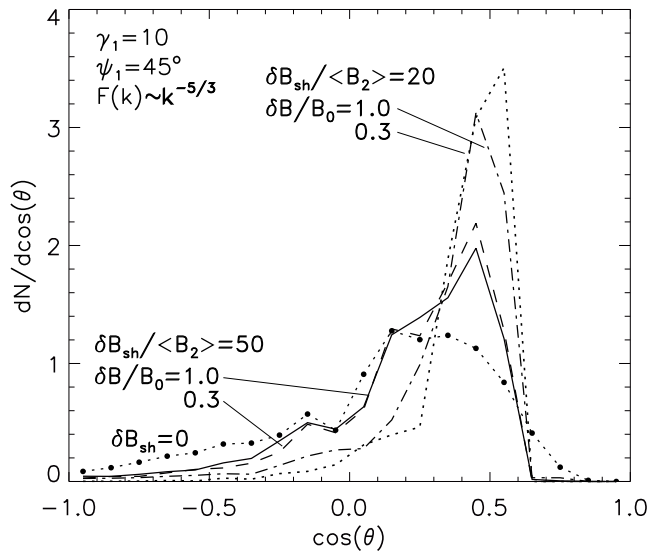}
\caption{Normalized particle angular distributions as measured in the shock 
rest frame for superluminal shock waves
with $\gamma_1=10$, $\psi_1=45^o$, and a Kolmogorov wave power spectrum of 
large-scale background magnetic field turbulence.
Particles with $\cos\theta<0$ are directed upstream of the shock.
The distributions are calculated by summing the quantity $w/(|v_x|+0.005)$ in 
the respective $\cos\theta$-bins at every particle shock crossing, 
where $\theta$ is the angle between
the particle momentum and the shock normal, $v_x$ is the normal component of the 
particle velocity, and $w$ is the statistical weight of the particle. Only particles with 
$E\geq 1$ are included, and low-energy particles with large
weights contribute most to the distributions.
The angular distribution presented with filled dots is derived
without short-wave perturbations ($\delta B_{sh}=0$). The distributions
for $\delta B_{sh}/\langle B_2\rangle=50$, and for $\delta B/B_{0}=1.0$ and 
$0.3$ are presented with solid and dashed lines, respectively. The angular
distributions for $\delta B_{sh}/\langle B_2\rangle=20$, and 
$\delta B/B_{0}=1.0$ and $0.3$ are shown with dash-dotted and dotted lines,
respectively.
\label{angsup}}
\end{figure} 

In the case of superluminal shocks, 
the particle spectra calculated for a high amplitude of
shock-generated turbulence, $\delta B_{sh}/\langle B_2\rangle=50$ in Figures
\ref{obl1}-\ref{obl4}, depend only weakly on the background large-scale magnetic
field structure, in contrast to the $\delta B_{sh} =0$ limit 
\citep[see also][]{nie06}. The acceleration process is only slightly more efficient
in the case of a large amplitude of long-wave perturbations 
($\delta B/B_{0}=1.0$ in Figs. \ref{obl1} and \ref{obl2}), in which the spectral
tails extend to marginally higher energies and show softer cutoffs, when compared 
to the cases with $\delta B/B_{0}=0.3$ (Figs. \ref{obl3} and \ref{obl4}).
There are also only minor differences between the spectra derived for a flat
spectrum of background perturbations, for which the wave power is 
uniformly distributed per logarithmic wavevector range, and for the Kolmogorov
distribution, for which most power is carried by long waves. In the latter case, 
one finds power-law particle spectra in a slightly wider energy range than 
in the case of flat distributions (compare Figs. \ref{obl1} and \ref{obl2} and 
spectra in Fig. \ref{obl4}), but the high-energy tails and the cutoff shape  
remain nearly the same for both types of the wave power spectra.

The structure of the background large-scale magnetic field can, however, 
influence particle acceleration for smaller amplitudes of the short-wave 
turbulence. As one can see in Figure \ref{obl2}, the power-law 
part of the spectrum for $\delta B_{sh}/\langle B_2\rangle=20$ and 
Kolmogorov-type long-wave turbulence spectrum, although steeper, continues 
to higher energies than that for $\delta B_{sh}/\langle B_2\rangle=50$,
which is caused by the large-scale component providing particle scattering at
higher energies (see, e.g., the spectrum for $\delta B_{sh} =0$ in Fig. 
\ref{obl2}). This is not the case for either a flat spectrum (Fig.~\ref{obl1}) 
or a small-amplitude (Fig.~\ref{obl3}) of long-wave turbulence, 
since in superluminal shocks particle acceleration processes in the absence of 
large-amplitude long-wave magnetic field perturbations are inefficient 
\citep{beg90,nie04,nie06}. 

Figure \ref{angsup} presents particle angular distributions for superluminal
shocks ($\psi_1=45^o$) with the Kolmogorov power spectrum of large-scale
magnetic field perturbations. Note that the angular distributions for a large 
amplitude of the small-scale turbulence, $\delta B_{sh}/\langle B_2\rangle=50$,
are closer to that for $\delta B_{sh}=0$ ({\it filled dots}) than angular 
distributions for the smaller amplitude, $\delta B_{sh}/\langle B_2\rangle=20$. 
As the particle spectra, the distributions for 
$\delta B_{sh}/\langle B_2\rangle=50$ do not depend on the amplitude of the 
background field perturbations, nor on their wave-power spectrum (distributions 
for the flat power spectrum are not shown for presentation
clarity). Similarity in shape of these distributions to the angular distribution 
derived without the $\delta B_{sh}$ scattering term is caused by efficient 
particle scattering on the downstream short-wave perturbations 
at low particle energies, which enables a considerable fraction of them to 
return upstream of the shock. 
The difference in the distributions for the smaller short-wave amplitude,
$\delta B_{sh}/\langle B_2\rangle=20$, illustrates the effects the long-wave
turbulence has on the particle acceleration process, as discussed above.
 
A caveat is in order concerning the low-energy part of the simulated spectra.
In our simulations with large-amplitude 
short-wave turbulence, the scattering amplitudes for low particle energies, 
$E\leq 1$ for the spectra with $\delta B_{sh}/\langle B_2\rangle=50$ and 
$E\leq 5$ for $\delta B_{sh}/\langle B_2\rangle=80$, violate the 
pitch-angle 
diffusion requirement $\Delta\Omega\ll 1$. This is because in our approach the
time between subsequent scatterings is fixed for a given particle energy, 
and the scattering angle scales linearly with $\delta B_{sh}$ (see \S 2.2) 
providing large scattering amplitudes for large values of $\langle B_2\rangle$. 
Therefore, the simulated spectra may suffer a weak systematic flattening
at low energies.

\subsection{Parallel Shocks}

\begin{figure*}
\begin{center}
\includegraphics[angle=0,scale=0.85]{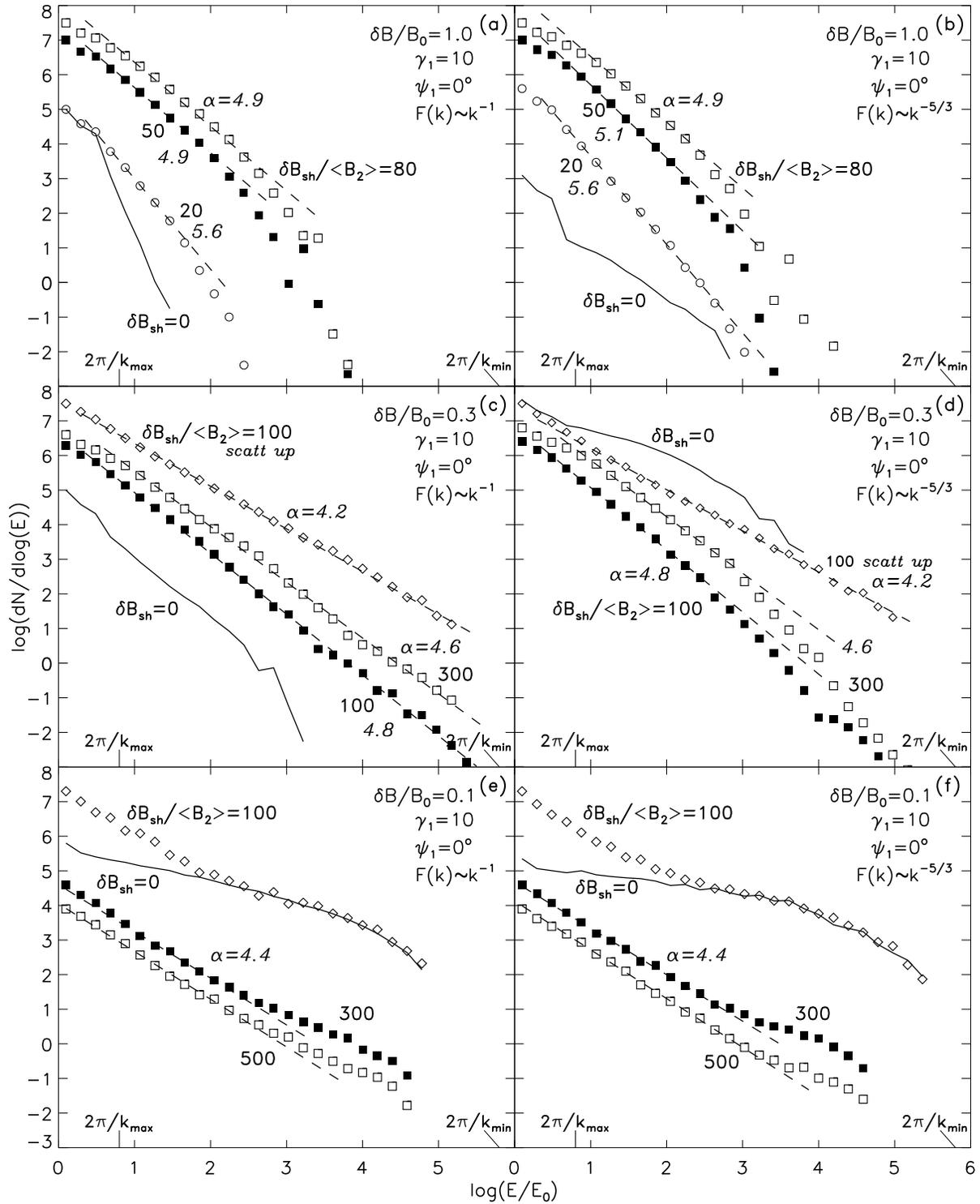}
\caption{Accelerated particle spectra at parallel ultrarelativistic shocks
($\gamma_1 = 10$, $\psi_1 = 0^o$). The parameters of the large-scale background 
magnetic field turbulence structure are provided in each panel. The spectra
shown with solid lines have been derived in simulations without small-scale
perturbations \citep[see Fig. 4 in][]{nie06}. The results indicated as 
``{\it scatt up}" in ({\it c}) and ({\it d}) show spectra obtained
in simulations for which the particle pitch-angle diffusion was
assumed to exist both downstream and upstream of the shock.  
\label{par10}}
\end{center}
\end{figure*}

\begin{figure*}
\begin{center}
\includegraphics[angle=0,scale=0.85]{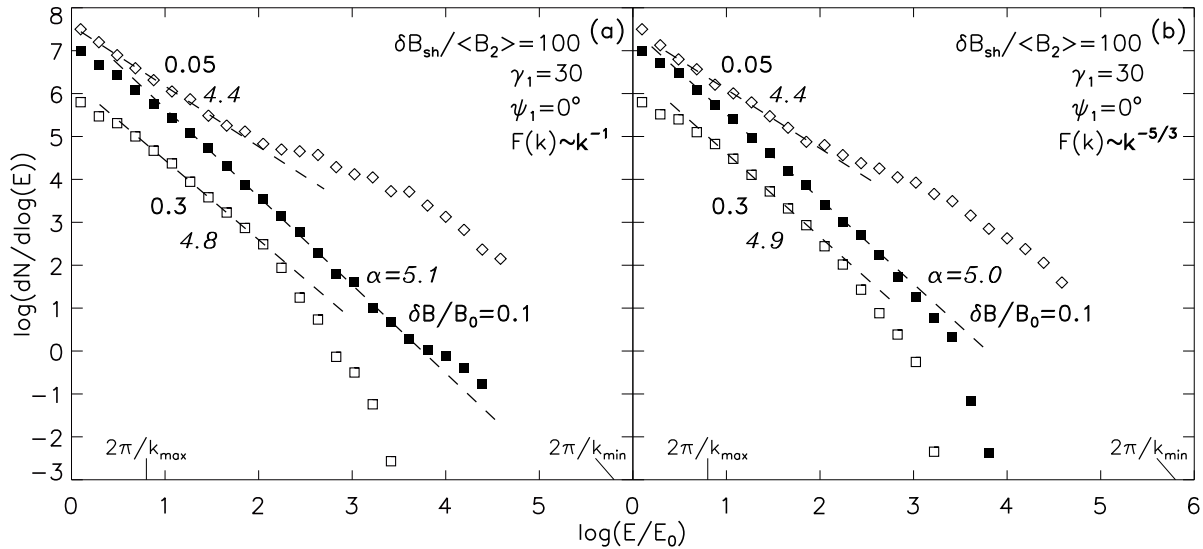}
\caption{Particle spectra at parallel ($\psi_1 = 0^o$) ultrarelativistic shocks
with $\gamma_1 = 30$, derived for a high amplitude of downstream short-wave
perturbations $\delta B_{sh}/\langle B_2\rangle=100$, and ({\it a}) the flat and
({\it b}) the Kolmogorov wave-power spectrum of large-scale background 
magnetic field turbulence. The amplitudes of the {\it background} perturbations 
$\delta B/B_{0}$ are given near the respective results, together with the values
of the spectral indices for fits to the power-law portions of the spectra.
\label{par30}}
\end{center}
\end{figure*}

As discussed by \citet{ost02}, the particle spectra with ``universal" 
slope have been derived in conditions equivalent to a parallel ($\psi_1=0^o$) 
mean shock 
configuration. In our earlier work \citep{nie04,nie06} we have demonstrated that 
the spectra formed at parallel relativistic shocks depend substantially on
background conditions, and features such as a relation between the spectral index 
and the turbulence amplitude or a cutoff formation in the case of ultrarelativistic
shocks can be observed. In this section, we investigate the role of  
shock-generated downstream turbulence in particle acceleration at parallel shocks. 
For that purpose, we have performed simulations for highly 
relativistic shock waves with Lorentz factor $\gamma_1=10$ and $30$, 
as presented in Figures \ref{par10} and \ref{par30}, respectively.
For parallel shocks the amplitude of the compressed background field 
downstream of the shock, $\langle B_2\rangle$, is on average much smaller than 
for the oblique shocks discussed in \S 3.1. Therefore, the numerical 
constraints are largely relaxed, and we can use larger values of 
$\delta B_{sh}/\langle B_2\rangle$, up to $300$ or even $500$. 
However, we would still slightly violate the pitch-angle scattering approximation
at $E\leq 1$ for the spectra with $\delta B_{sh}/\langle B_2\rangle=80$ in Figs.
\ref{par10}{\it a} and \ref{par10}{\it b}, and 
$\delta B_{sh}/\langle B_2\rangle=300$ in Figs. \ref{par10}{\it c} and 
\ref{par10}{\it d}.

Figures \ref{par10}{\it a} and \ref{par10}{\it b} show the results for shocks
with $\gamma_1 = 10$ and large-amplitude background field perturbations
($\delta B/B_{0} = 1.0$). The qualitative similarity of the spectral features
observed in this case and for oblique superluminal shocks is apparent 
\citep[see also][]{nie06}---the spectra are convex and display
cutoffs occurring much below $E_{res,max}\simeq 2\pi/ k_{min}$. 
This behavior proves that the large-amplitude long-wave perturbations locally form 
superluminal conditions at the shock, thus leading to a spectral cutoff
as the pitch-angle scattering term in the 
downstream shock-generated turbulence substantially decreases with growing 
particle energy. Note, that in the case of a lower amplitude of the short-wave
component, $\delta B_{sh}/\langle B_2\rangle=20$, the large-scale background
turbulence with Kolmogorov distribution provides additional particle scattering, 
allowing for the formation of a quasi--power-law spectrum in a slightly wider 
energy range 
than for the larger $\delta B_{sh}/\langle B_2\rangle$ (Fig. \ref{par10}{\it b}),
again in correspondence to the analogous result for the superluminal shocks
(Fig. \ref{obl2}).

\begin{figure}
\includegraphics[angle=0,scale=0.85]{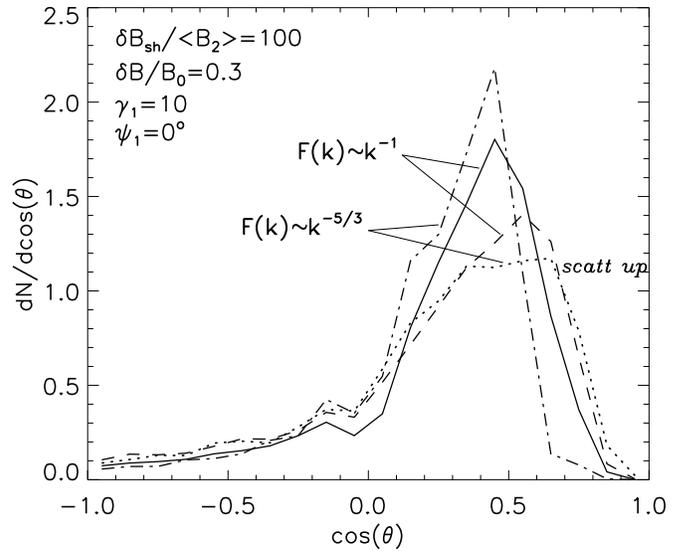}
\caption{Particle angular distributions at parallel shocks ($\gamma_1 = 10$, 
$\psi_1 = 0^o$) for $\delta B_{sh}/\langle B_2\rangle=100$ and a weakly perturbed
large-scale background magnetic field ($\delta B/B_{0}=0.3$) with either the 
flat (solid and dashed line) or the Kolmogorov (dash-dotted and dotted line) 
wave power spectrum of magnetic field perturbations. The distributions indicated as
"{\it scatt up}" are derived in a model with short-wave turbulence imposed both
downstream and upstream of the shock.
\label{angpar}}
\end{figure}

For the smaller amplitude of the large-scale magnetic field 
perturbations, $\delta B/B_{0}=0.3$, the role of the long-wave 
perturbations is less significant 
and the scattering on the short-wave turbulence can dominate up to the 
energies higher than the upper resonance energy $E_{res,max}$. 
Though being gradually weaker with growing particle
energy, in this case, as shown in Figure \ref{par10}{\it c}, pitch-angle diffusion 
can dominate over scattering by the long-wave perturbations without a 
limit for particle energy below 
$E_{res,max}$. The scattering mean 
free paths along the mean background magnetic field (shock normal) and, 
correspondingly, the mean acceleration timescale also increase with particle energy,
but the particle spectra retain a power-law form up to the 
highest energies $E\approx E_{res,max}$  studied in our simulations. 
However, the spectra are steeper than the expected ``universal'' spectrum,
$\alpha > \alpha_u$, and only weakly dependent on $\delta B_{sh}$. 
To understand this feature, we have performed additional test simulations with 
the short-wave component introduced both downstream and {\it upstream} of the 
shock (with constant upstream scattering amplitude 
$\Delta\Omega(E)_{up}=const\ll 1/\gamma_1$). This 
case, equivalent to the pure pitch-angle diffusion model, should lead to the 
formation of the ``universal" spectrum and, indeed, such the spectrum is 
produced in our simulations
(described as ``{\it scatt up}" in Fig. \ref{par10}{\it c}). Thus, 
the slightly steeper spectra obtained in the models with only a
downstream short-wave component can be understood solely as the result of the
upstream long-wave turbulence providing a different distribution of pitch-angle 
perturbations to the particle orbits than the purely diffusive process
(see Fig. \ref{angpar}). This allows for the generation of
a power-law particle spectrum, but with somewhat steeper index 
than the ``universal" value. 

With the Kolmogorov background wave spectrum (Fig. \ref{par10}{\it d}) there is 
enough power in long waves to influence the spectrum at high energies,
leading to the noticable steepening of the particle distributions below  
$E_{res,max}$. However, also in this case the model with the short-wave 
field perturbations imposed upstream of the shock yields the ``universal" 
power-law spectrum. In Figure \ref{angpar} we compare
the angular distributions of particles at the shock in the cases of 
only downstream small-scale turbulence and with 
pitch-angle diffusion also upstream of the shock.
  
The spectra for weakly perturbed background magnetic fields, 
$\delta B/B_{0}=0.1$, presented in Figures \ref{par10}{\it e} and 
\ref{par10}{\it f}, show a new characteristic feature, namely a spectral 
bump besides the power-law spectrum at lower particle 
energies. The bump shape closely resembles the distribution of particles accelerated 
without the shock-generated field component 
\citep[see the discussion in][]{nie06}, and its presence indicates that
at these energies the short-wave perturbations become
negligible compared to the background long-wave component. 
Note, that $\delta B_{sh}$ is scaled with respect to $\langle B_2\rangle$,
which is relatively small in comparison to the cases with larger 
$\delta B/B_{0}$ discussed above.

We have also performed a series of simulations for shocks with a larger Lorentz
factor of $\gamma_1 = 30$. The particle spectra
presented in Figure \ref{par30} for $\delta B_{sh}/\langle B_2\rangle=100$ show 
a similarity of spectral features to those observed for shocks with 
$\gamma_1 = 10$. However, there is an approximate scaling between the two cases, 
with characteristic spectral features appearing at respectively lower 
amplitudes of the long-wave magnetic field 
perturbations $\delta B/B_{0}$ for shocks with $\gamma_1 = 30$ 
than in the spectra for $\gamma_1 = 10$. 
Such a scaling may result from the stronger compression of these perturbations, 
as measured between the two plasma rest frames, and also from the larger particle 
anisotropy involved at the higher-$\gamma$ shock. 

\section{CONCLUDING REMARKS}
The present paper concludes our studies \citep{nie04,nie06} of the first-order 
Fermi acceleration mechanism acting at relativistic and ultrarelativistic shock 
waves. We have attempted to consider as realistic models as possible 
for the perturbed magnetic field structures at the shock, which allow us to study 
all the field characteristics important for particle acceleration. 
In particular, we have investigated the dependence of  particle spectra on  
the mean magnetic field inclination with respect to the shock normal and 
on the power spectra of
magnetic field perturbations, including the long-wave background 
turbulence and the short-wave turbulence generated at the shock. 

In the present  paper we append the model of \citet{nie06} with downstream 
large-amplitude small-scale MHD turbulence component, 
analogous to those seen in PIC simulations of 
collisionless relativistic shocks. The form of these additional perturbations is 
arbitrarily chosen to consist of a very short sinusoidal waves, that form
(static) isotropic highly nonlinear turbulence. We have considered a wide variety of 
ultrarelativistic shock configurations for which, as presented in \S 3, 
a number of different spectral features can be observed. We show that with 
growing particle energy the role of short-wave ($\lambda_{sh}\ll r_g(E)$) 
magnetic field perturbations decreases, and spectra at high energies are shaped 
only by the mean magnetic field and the long-wave 
perturbations. Thus, in oblique superluminal shocks concave particle spectra with
cutoffs can be generated even for $\delta B_{sh}\gg\langle B_2\rangle$. 
Extended power-law particle distributions can be formed in parallel shocks 
propagating in a medium with low-amplitude long-wave field perturbations, but 
the spectral indices obtained
are larger than the ``universal" spectral index 
$\alpha_u\approx 4.2$, that is widely considered in the literature. The only case in
which we have been able to obtain spectra with $\alpha = \alpha_u$  involved the
unphysical assumption, that large-amplitude short-wave field perturbations also 
exist upstream of the shock. 

In summary, the accelerated particle spectral distributions obtained in this work
and in our previous studies \citep{nie04,nie06} generally differ from the spectra
of relativistic electrons (power-laws with the spectral indices close to 
$\alpha_u$) inferred from modeling the electromagnetic emission spectra of astrophysical
sources hosting relativistic shocks (e.g., hot spots in extragalactic radio sources, quasar 
jets, gamma-ray burst afterglows). Our results thus provide a strong argument 
against considering the first-order Fermi shock acceleration as the main 
mechanism producing the observed radiating electrons. In our opinion, other 
processes must be invoked 
to explain the observed emission spectra, for example second-order Fermi 
processes acting in the regions of relativistic MHD turbulence downstream of the
shock, which have hardly ever been discussed for relativistic 
conditions till now \citep{vir05}, or the collisionless plasma processes 
that have been studied in numerous PIC simulations
\citep[e.g.,][]{hos92,hed04,nis05b}, or other non-standard acceleration 
processes like those discussed by 
\citet{ste03} for electrons or \citet{der03} for both electrons and protons. 

Our simulations also show, 
that relativistic shocks, being essentially always superluminal, 
possibly generate accelerated particle distributions with cutoffs below either 
the maximum resonance energy enabled by the {\it high-amplitude} background 
turbulence ($r_g(E_{cutoff}) < \lambda(E_{res,max})$), or approximately at the 
energy of the compressed background plasma ions 
$E_{cutoff} \sim \gamma_1 m_{i}c^2$ (where $m_{i}$ is a mass of the heaviest 
ions present in the background medium). Thus, in conclusion, we maintain our 
opinion from the previous publications \citep[see also][]{beg90} that 
relativistic shocks are not promising sites as possible sources of  
ultra--high-energy nuclei registered by the air shower experiments. 

\acknowledgments
We are grateful to Mikhail Medvedev for interesting discussions. The present 
work was supported by MNiI in years 2005-2008 as a research project 
1 P03D 003 29.

\end{document}